\documentclass{article}
\usepackage{spconf,amsmath,graphicx}
\usepackage {sub figure} 

\usepackage{enumitem}
\setlist{nosep, leftmargin=14pt}

\usepackage{mwe} % to get dummy images

\title{Weakly Supervised Learning for cell recognition in immunohistochemical cytoplasm staining images}
\name{
\begin{tabular}{@{}c@{}}
Shichuan Zhang$^{1,2,3}$ \qquad Chenglu Zhu$^{1,2}$ \qquad Honglin Li$^{1,2}$ \qquad Jiatong Cai$^{1,2}$  \qquad Lin Yang$^{1,2}$ 
\end{tabular}
}

\address{$^{1}$Artificial Intelligence and Biomedical Image Analysis Lab, School of Engineering, Westlake University\\
    $^{2}$Institute of Advanced Technology, Westlake Institute for Advanced Study\\
    $^{3}$College of Computer Science \& Technology, Zhejiang University
    }
\begin{document}
%\ninept
%
\maketitle
\begin{abstract}
% Cell classification and counting in immunohistochemical cytoplasm staining images play a pivotal role in cancer diagnosis. To avoid the labor-intensive labeling work, we design a multi-task learning algorithm based on point annotation. Considering the unique challenges for cell recognition in immunohistochemical cytoplasm staining images, two auxiliary branches are appended to explicitly provide more information for the features extractor (encoder). For misclassification of some cells, the auxiliary branch (prior decoder) provides spatial distribution as an additional cell classification basis. For the missed detection due to differences in cell morphology, consistency learning between the output of main branch (main decoder) and the auxiliary branch (dynamic decoder) is used to avoid overfitting the proximity masks. We have evaluated our framework on immunohistochemical cytoplasm staining images, and the experiment results illustrate that our method outperforms recent cell recognition approaches. Besides, we have also done some ablation studies to show significant improvement after adding the auxiliary branches.
Cell classification and counting in immunohistochemical cytoplasm staining images play a pivotal role in cancer diagnosis. Weakly supervised learning is a potential method to deal with labor-intensive labeling. However, the inconstant cell morphology and subtle differences between classes also bring challenges. To this end, we present a novel cell recognition framework based on multi-task learning, which utilizes two additional auxiliary tasks to guide robust representation learning of the main task. 
To deal with misclassification, the tissue prior learning branch is introduced to capture the spatial representation of tumor cells without additional tissue annotation. Moreover, dynamic masks and consistency learning are adopted to learn the invariance of cell scale and shape.
We have evaluated our framework on immunohistochemical cytoplasm staining images, and the results demonstrate that our method outperforms recent cell recognition approaches. Besides, we have also done some ablation studies to show significant improvements after adding the auxiliary branches.
\end{abstract}

\section{INTRODUCTION}
% Quantitative analysis of pathological images is a vital step in cancer diagnosis, prognosis, and treatment decisions. Automatic cell recognition (classification and counting) algorithms\cite{qu2020weakly,xie2018efficient} based on nucleus staining reduce the intensive labor and improve the efficiency for pathologists.  Immunohistochemical (IHC) cytoplasm staining images also play an important role for the pathologist to ascertain the positive-tumor cells\cite{weidemann2021napsin}. And cells recognition is useful for confirming the type of cancer, stage of cancer, and potential treatment options based on IHC cytoplasm staining images\cite{zhou2010mtor}. Therefore, the automatic analysis algorithm based on deep learning for IHC cytoplasm staining images can help clinical diagnosis.
Immunohistochemical (IHC) staining is a universal protocol to signalize specific tumor cells. The cancer progress can be evaluated by quantitating the response of different cells to antigens, which promotes making prognostic observation and treatment decisions. 
Recent automatic cell recognition (classification and counting) algorithms~\cite{qu2020weakly,xie2018efficient} based on nucleus staining reduce the labor-intensive counting and improve the efficiency of pathologists. 
Some of the specific IHC antibody reactions occur in the cytoplasm~\cite{weidemann2021napsin}, which will bring some challenges to current IHC analysis models with their unclear boundaries and confusing nucleus. 
Therefore, it is significant to propose a general analysis method based on deep learning for IHC cytoplasm staining images, which can help to confirm the type of cancer, stage of cancer, and potential treatment options in clinical diagnosis~\cite{zhou2010mtor}.

% Recently, there are many works on nuclear recognition that have shown superior performance which are based on point annotation with weakly supervised. The authors regard an gaussian density map from the point annotations as the estimated ground truth mask in \cite{xie2018microscopy}. Then the regression loss is computed for training. And the centers of nucleus closed to a circle or ellipse is paid attention to. In this paper, we focus on the cells recognition in IHC cytoplasm staining images, in which the cytoplasm is expressed and cells are in various shapes. So the regression method may not perform as well as nuclear recognition. A pixel classification method is also used the for cell recognition\cite{qu2020weakly}. The author uses a relatively novel way to generate pseudo masks that are as close to the ground truth masks as possible. There is a fine tuning step based on conditional random fields (CRF) after the training with pseudo masks. For IHC cytoplasm staining cells, the features are shown in the cytoplasm, which affects the generation of the pseudo mask based on such clustering algorithm\cite{qu2020weakly}. Besides, small differences between tissue cells and positive-tumor cells caused by the immunohistochemistry process are inevitable. All these problems bring new challenges for cell recognition in the cytoplasm staining images.
Recently, many nuclear recognition models with regression loss for IHC image analysis have shown superior performance, which applies weakly supervised learning under the generated Gaussian density map from point annotations~\cite{xie2018microscopy}.
However, the cytoplasmic staining pattern does not have uniform recognition objects, and its stained images are different from the nuclear-stain with stainable antibodies in the circle or ellipse nucleus. After cytoplasmic staining, cells may take on various shapes and the neighbors may present indistinguishable masses. Consequently, it may not be appropriate to use the general regression models for nuclear recognition in this scenario.
Besides, a pixel classification method~\cite{chamanzar2020weakly} for cell recognition improves model performance by multi-task scheduling, which uses repel code, Voronoi diagram and clustering algorithms to strengthen the supervision information of point annotation.
% Besides, a pixel classification method~\cite{chamanzar2020weakly} is also used for cell recognition, which uses repel code, Voronoi diagram and clustering algorithms to strengthen the supervision information of point annotation, and collaboratively improves model performance by multi-task scheduling.
% In addition, the fine tuning step is based on conditional random fields (CRF) after the training with pseudo masks.
Nevertheless, cells are touching and overlapping each other under cytoplasmic staining patterns, which is not conducive to generating the pseudo mask with such clustering algorithms~\cite{qu2020weakly}.
In addition, subtle differences between tissue cells and positive-tumor cells are inevitable in IHC images,  which increases the difficulty of categorization. All these problems bring new challenges for cell recognition in the cytoplasm staining images.
% Nevertheless, it is difficult to distinguish neighboring cells under cytoplasmic staining patterns, which is not conducive to generate the pseudo mask with the clustering algorithm~\cite{qu2020weakly}.
% In addition, inevitable subtle differences between tissue cells and positive-tumor cells in the IHC process may increase the difficulty of categorization. All these problems bring new challenges for cell recognition in the cytoplasm staining images.

% Multi-Task Learning(MTL) aims to improve the generalization of the model by leveraging the training information in the main task and the auxiliary tasks\cite{2020Multi}. MTL has been applied to various downstream tasks including cell recognition. The authors in \cite{2019Pixel} designed an auxiliary task to provide useful tissue prior information for the encoder. The spatial distribution information from the tissue prior is also the basis for cell classification. But they need additional tumor area labeling. 
% The estimated mask is a commonly used method for cell recognition with point annotation. But the model can easily overfit in similar shape and the same size of each cell. Consistency learning is an effective regularization strategy to improve the performance. \cite{2021Dual} and \cite{2020Semi} both append auxiliary decoders guided by uncertainty masks for semi-supervised semantic segmentation, and then minimize the consistency loss of the output from different branches. Thus, the rough pseudo masks can be used effectively.

\begin{figure*}[htbp]
\centering
\includegraphics[width=1.9\columnwidth]{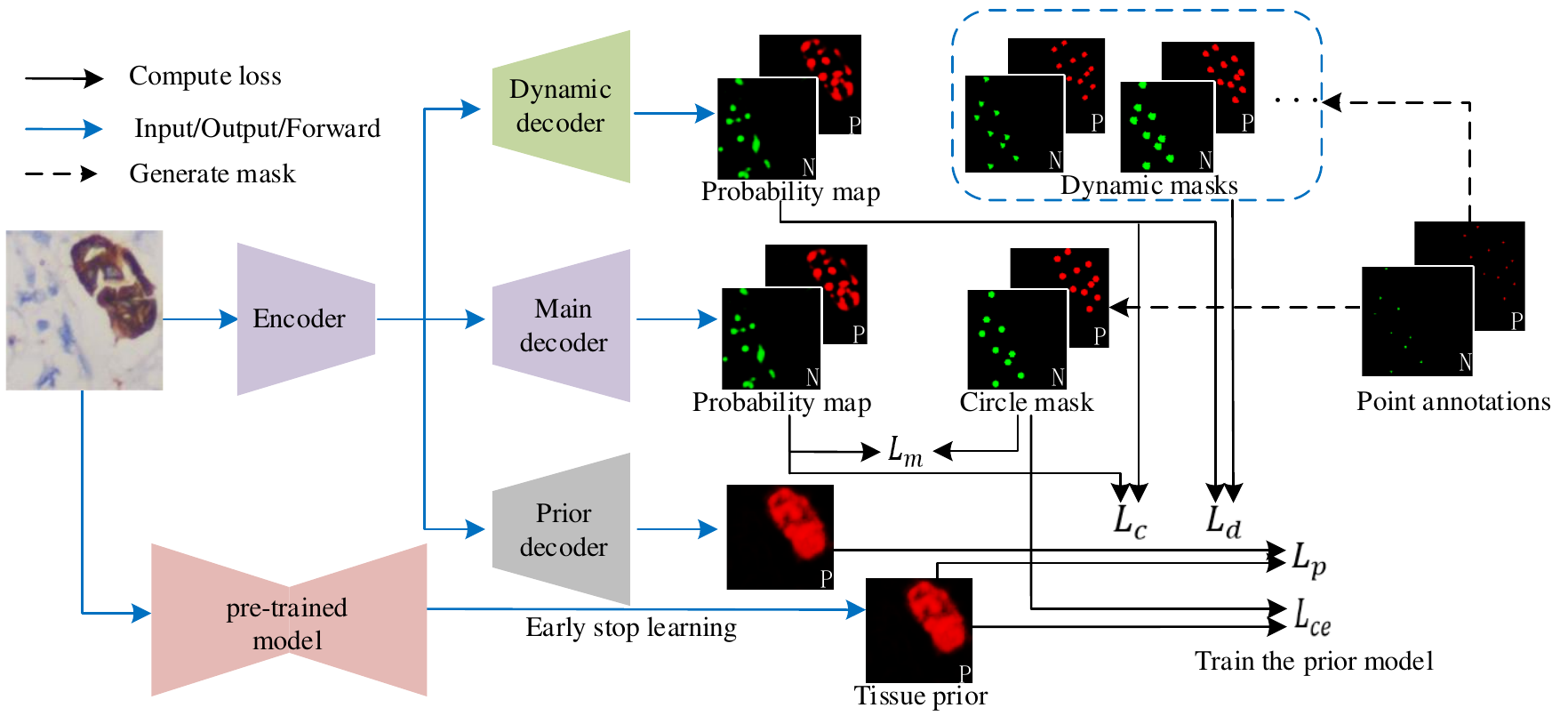} 
% \vspace{-0.3cm}
\caption{The whole framework based on multi-task learning. Three branches learn their own tasks by $L_m$, $L_d$ and $L_p$ respectively. Dynamic masks and circle masks are generated from point annotations. The prior masks are the output of a pre-trained model learned by $L_{ce}$. The consistency loss between the output of main decoder and dynamic decoder is minimized by $L_c$.}
\label{fig:overview}
\end{figure*}

The generated pseudo masks from point annotations are used for model regression. Due to the limitation of Gaussian field, the model may easily overfit to a fixed size and approximately round shape, which may become more serious in the cytoplasmic staining pattern. Fortunately, consistency learning is an effective regularization strategy to improve the availability of rough pseudo masks. It minimizes the output of each auxiliary decoding branch through consistency loss in the semi-supervised semantic segmentation tasks~\cite{li2021dual,2020Semi}, which promotes model representation learning based on uncertain masks. Moreover, tissue prior can guide cell recognition by introducing an auxiliary task~\cite{xing2019pixel}, which is constructed to learn the spatial distribution of tumors with extra annotations.
A traditional multi-task learning framework whose auxiliary tasks are to learn from different types of pseudo masks has been introduced to deal with nuclear recognition. So well-designed auxiliary tasks can effectively improve the generalization of the main task.
% Multi-task Learning has been introduced to deal with difficult  classification~\cite{2020Multi}, which leverages the training information between the main task and auxiliary tasks to improve the model generalization.
% The tissue prior can constrain the cell recognition by introducing an auxiliary task~\cite{2019Pixel}, which is constructed to learn the spatial distribution of tumors with extra annotations. 

% Inspired by multi-task learning and consistency learning, we design a framework with three branches: main decoder, dynamic decoder and prior decoder. Another model trained with early stop is used for generating the tissue prior. Therefore, the encoder and the auxiliary branch(prior decoder) can be guided by the tissue prior even without tumor area labeling. Prior decoder provides additional spatial basis for cell classification. The other auxiliary branch is guided by a dynamic mask in which the marks of cells are with different shapes and sizes generated from point annotation randomly. Last, we minimize the consistency loss between the outputs of the main decoder and the dynamic decoder to avoid overfitting. 
Inspired by explicit tissue prior and consistency learning, we present a cell recognition framework for the quantification of IHC cytoplasmic staining images by introducing three decoder branches: main decoder, dynamic decoder and prior decoder. 
Since the tissue prior is generated by an additional model with early stop learning, the encoder and the auxiliary branch(prior decoder) can be constrained by spatial distribution information without extra tissue annotations.
%%
% In addition, dynamic supervisied learning is adopted in the other auxiliary branch to generalize the model representation by the stochastic disturbance of generates masks with different shapes and sizes.
% % Last, we minimize the consistency loss between the outputs of the main decoder and the dynamic decoder to avoid overfitting.
% The other auxiliary branch(dynamic decoder) and the encoder are guided by dynamic masks in which the marks of cells are with different shapes and sizes. 
In addition, the other auxiliary branch (dynamic decoder) and the encoder are guided by dynamic masks with stochastic disturbance. Last, the consistency loss between the outputs of the main decoder and the dynamic decoder is minimized to learn the invariance of shape and scale. 

% The encoder and the main decoder are used for testing. Dynamic decoder and prior decoder offer more information of the data for the encoder during training. 

% 1. Hard or soft mask(classification or regression) for counting task.

% 2. Cell counting nucleus and membrane

% 3. Consistent learning of counting task

% 4. Organizational prior

% Challenge--

% Contribution--

\section{METHOD}

The proposed framework is shown in Fig.\ref{fig:overview}. It contains three parts: encoder, three decoders and a pre-trained model. 
And the decoder consists of three branches: main branch, consistency learning branch and tissue prior learning branch. 
In the whole framework, only the encoder and the main decoder exist in both training and testing processes. While the other two auxiliary branches provide more information to the encoder during training. The encoder downsampling the input image to a deep feature and the decoders upsampling the feature to the original size construct the forward process.

\subsection{Main branch}
Semantic segmentation based on an encoder-decoder structure is used for cell counting and classification. 
% We have mentioned that it is inappropriate to use the gaussian density maps based on point annotation for the recognition of cytoplasmic-stained cells. 
Thus, a fixed-size circle mark is generated for each cell during the training of the main decoder. The estimated shape and size are chosen to approximate the ground truth mask for cell segmentation and classification. There are two proximity masks $\{G_{i,j}^l|G_{i,j}^l\in \{0,1\},0<i<m,0<j<n\}$ for each input image with size $m\times n$, which one is for positive-tumor cells and the other one is for the rest cells. The main decoder outputs two corresponding probability map $\{P_{i,j}^l| 0<P_{i,j}^l<1,0<i<m,0<j<n\}$. $l\in\{0,1\}$ represents the category of cells, $(i,j)$ is the pixel coordinates. A certain value $P_{i,j}^l$ in the output map is the probability that the corresponding input pixel belongs to class $l$. Thus we use the cross-entropy loss for pixels classification as shown in eq.\ref{eq:BCE}
% \vspace{-1em}
\begin{equation}
    L_{ce}=\frac{1}{2mn}\sum ^{1}_{l=0}\sum ^{m}_{i=1}\sum ^{n}_{j=1}-G_{i,j}^llogP_{i,j}^l
\label{eq:BCE}
\end{equation}
Taking into account the imbalance between the number of object pixels and the number of background pixels, we also apply the Intersection over Union(IOU) loss\cite{yu2016unitbox} for the training of the main decoder. And the IOU loss as shown in eq.\ref{eq:IOU} is sensitive to the imbalance.
\begin{equation}
    L_{IOU}=\frac{1}{2}\sum ^{1}_{l=0}\frac{\sum ^{m}_{i=1}\sum ^{n}_{j=1}G_{i,j}^l*P_{i,j}^l}{\sum ^{m}_{i=1}\sum ^{n}_{j=1}G_{i,j}^l+\sum ^{m}_{i=1}\sum ^{n}_{j=1}P_{i,j}^l}
\label{eq:IOU}
\end{equation}
\begin{equation}
    L_{m}=\alpha L_{ce} +(1-\alpha)L_{IOU}
\label{eq:LM}
\end{equation}
Eq. \ref{eq:LM} shows the total loss for the training of the main decoder and the encoder. $\alpha$ is a hyperparameter whose value is $0.8$ in our experiments.

\subsection{Auxiliary branch--consistency learning}
Not all cells are approximately round, especially in IHC images with cytoplasmic staining. The cells usually have different sizes and shapes, even in the same category. So the encoder and the main decoder trained with $L_m$ is easy to cause overfitting and missed detection. 
% Inspired by the consistency learning in semi-supervised or unsupervised classification tasks\cite{2021Dual}, which minimize the prediction difference of the same image after different transformations. 
In order to make the model not sensitive to inaccurate shapes and sizes when extracting features, We design an auxiliary branch (dynamic decoder). In each training iteration, the cell marks with random polygon and size in new estimated masks are generated for the training of the encoder and the dynamic decoder by eq.\ref{eq:dynamic_loss}. Then we minimize the difference of the output probability maps between dynamic decoder and main decoder by the consistency loss as shown in eq.\ref{eq:consistent}.
\begin{equation}
\setlength{\abovedisplayskip}{5pt}
\setlength{\belowdisplayskip}{5pt}
    L_{d}=\frac{1}{2mn}\sum_{l=0}^{1}\sum_{i=1}^{m}\sum_{j=1}^{n}||\Tilde{P}_{i,j}^l-\Tilde{G}_{i,j}^l||,
\label{eq:dynamic_loss}
\end{equation}
\begin{equation}
\setlength{\abovedisplayskip}{5pt}
\setlength{\belowdisplayskip}{5pt}
    L_{c}=\frac{1}{2mn}\sum_{l=0}^{1}\sum_{i=1}^{m}\sum_{j=1}^{n}||P_{i,j}^l-\Tilde{P}_{i,j}^l||,
\label{eq:consistent}
\end{equation}
where $\Tilde{G}_{i,j}^l$ is the generated dynamic masks in which the proximity marks of cells have distinct shapes and sizes in different iterations. And $\Tilde{P}_{i,j}^l$ is the output probability map of the dynamic decoder. By consistency learning, the lower layers in the encoder will focus on the features that are conducive to cell identification in addition to size and shape. It is an effective regularization strategy for the encoder, which reduces the risk of overfitting the proximity masks.

\subsection{Auxiliary branch--tissue prior}
In the IHC images, the same kind of cells are generally distributed in clusters. In other words, tumor cells are usually adjacent to tumor cells, and vice versa. This \emph{\textbf{tissue prior}} is helpful to classify different types of cells with similar appearance characteristics. In practice, the pathologists identify tumor cells based on not only cell appearances, such as size, shape and intensity, but also on the spatial distribution of cells. Thus we design another auxiliary branch (prior decoder). It has the same layers as the dynamic decoder and main decoder. Positive-tumor cells are generally arranged tightly and touching to each other, so the combination of the rough area of each positive-tumor cell will form the proximity mask for the tumor area. The ground truth mask for the training of prior decoder and encoder is from the output of a pre-trained model. The pre-trained model has the same architecture as the combination of the encoder and the main decoder and is trained by the circle masks. In circle masks, the mark of each cell is a circle with a specified radius, resulting in inaccurate labeling of cell boundaries. Inspired by the robust learning for classification with label noise\cite{arazo2019unsupervised}, an early stop training process with $L_{ce}$ is used for the pre-trained model. And the output will show the rough tissue prior as shown in Fig.\ref{fig:overview}. The training loss for the prior decoder is shown in eq.\ref{eq:prior_loss}.
% \vspace{-1em}
\begin{equation}
\setlength{\abovedisplayskip}{5pt}
\setlength{\belowdisplayskip}{5pt}
    L_{p}=\sum_{i=1}^{m}\sum_{j=1}^{n}||\hat{G}_{i,j}^l-\hat{P}_{i,j}^l||, l=1, 
\label{eq:prior_loss}
\end{equation}
where $\hat{G}_{i,j}^l$ is the ground truth mask for the training of prior decoder and encoder from the output of a pre-trained model. And $\hat{P}_{i,j}^l$ is the output of prior decoder. Only positive-tumor prior is used in this branch because other types of tissue cells are not in a state of aggregation. All the three decoders corresponding to three tasks share the same encoder in the lower layers. The two auxiliary tasks make the encoder pay attention to their own tasks which are useful for the main task--cell classification and counting. Therefore, the encoder will benefit from the two auxiliary tasks in representation learning.

% \vspace{-0.5cm}
\section{EXPERIMENTS}
\subsection{Datasets and implementation details}
% We use IHC cytoplasmic staining images for model training and testing. The dataset includes 80 1920×1080 patches. And all this pathces are cut from $\times40$ magnification whole slide images. The cells are divided into two categories: positive-tumor cells (dark or weak brownish stains) and non-positive-tumor cells. We split the dataset into training($70\%$) and testing set($30\%$). And both the training set and testing set has all types of cells, covering IHC images of four types of cancer stained with different antibodies. The proportion of positive-tumor cells plays an important role in the quantitative diagnosis of all the types of cancer\cite{weidemann2021napsin,zhou2010mtor}. We count the total number of different types of cells in the training set and test set as shown in table \ref{tab:data}.
We collected 80 images with IHC cytoplasmic staining cropped as the heat map from whole slide images under×40 magnification, which cover four types of cancer and their corresponding stained antibodies. All cells are separated into two categories by two pathologists: positive tumor cells (dark or weak brownish stains) and the other cells without IHC reaction. 
% We split the dataset into training set ($70\%$) and testing set ($30\%$) contains all types of cancer cells. The detail of dataset is shown in table \ref{tab:data}.
We split the dataset into a training set (56 images with annotated 15,309 positive tumor cells / 37,277 other cells) and a testing set (24 images with annotated 6,053 positive tumor cells / 8,821 other cells). 
% \begin{table}
% \begin{center}
% \caption[]{Details of the training and testing data}
% \label{tab:data}
% \resizebox{\columnwidth}{!}{
% \begin{tabular}{ccccc}
% \hline\noalign{\smallskip}
% Dataset & Number of Images & Size& Positive tumor cells & Other cells \\
% \hline\noalign{\smallskip}
% $train$ &$56$ &$1920\times1080$ &$15309$ &$37277$\\
% $test$ &24 &$1920\times1080$ &$6053$ &$8821$\\
% \noalign{\smallskip}\hline
% \end{tabular}}
% \end{center}
% \end{table}
% In the whole framework as shown in Fig.\ref{fig:overview}, the encoder and main decoder follow the structure of the commonly used segmentation model deeplabv3+\cite{chen2018encoder} with resnet101 backbone, which contains less parameters than Unet\cite{ronneberger2015u}. The dynamic decoder and prior decoder has same layers to the main decoder. We first train the pre-trained model with early stop to get the rough cluster prior for tumor cells. And then use the total loss computed from the three branches to train the model. The total loss is shown in eq.\ref{eq:total_loss}
In the whole framework as shown in Fig.\ref{fig:overview}, the encoder and main decoder follow the structure of the commonly used segmentation model deeplabv3+~\cite{chen2018encoder} with resnet101 backbone, which has fewer parameters than Unet~\cite{ronneberger2015u}. 
The dynamic decoder and prior decoder in the proposed framework have the same layers as the main decoder. 
We first train the pre-trained model with early stop in 10 epochs to get the rough cluster prior for tumor cells. And then use the total loss computed from the three branches to train the model. 
\begin{equation}
    L_{t}=L_{m} +\lambda_c L_{c}+\lambda_p L_{p}+\lambda_d L_{d}, 
\label{eq:total_loss}
\end{equation}

\begin{figure*}
\centering
% \vspace{-0.3cm}
\subfigure[$Ours$]{
\includegraphics[width=0.6\columnwidth,height=0.55\columnwidth]{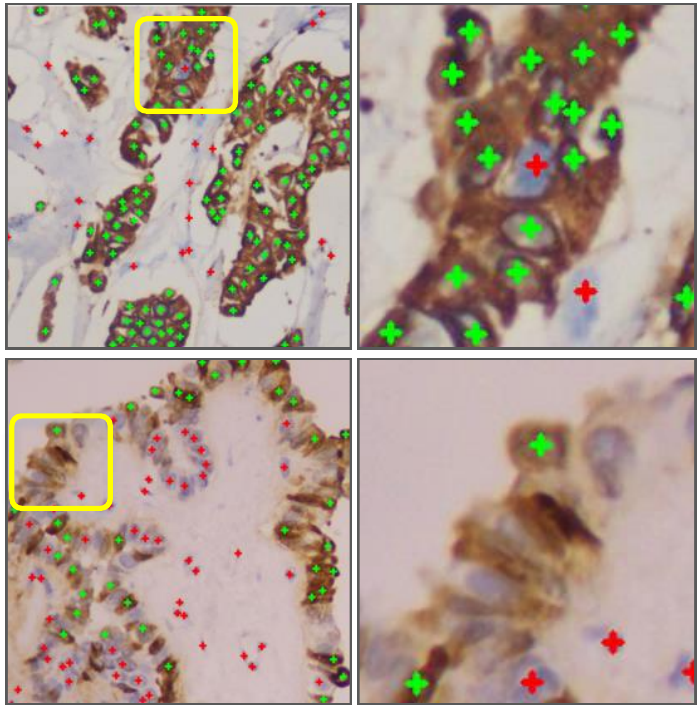}
}\hspace{-1mm}
\subfigure[$Ours^{++}$]{
\includegraphics[width=0.6\columnwidth,height=0.55\columnwidth]{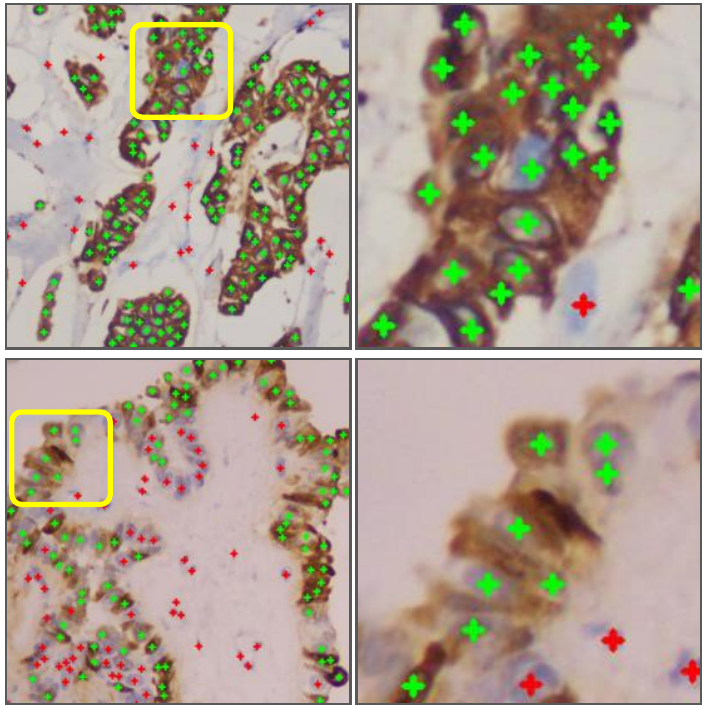}
}\hspace{-1mm}
\subfigure[Ground Truth]{
\includegraphics[width=0.6\columnwidth,height=0.55\columnwidth]{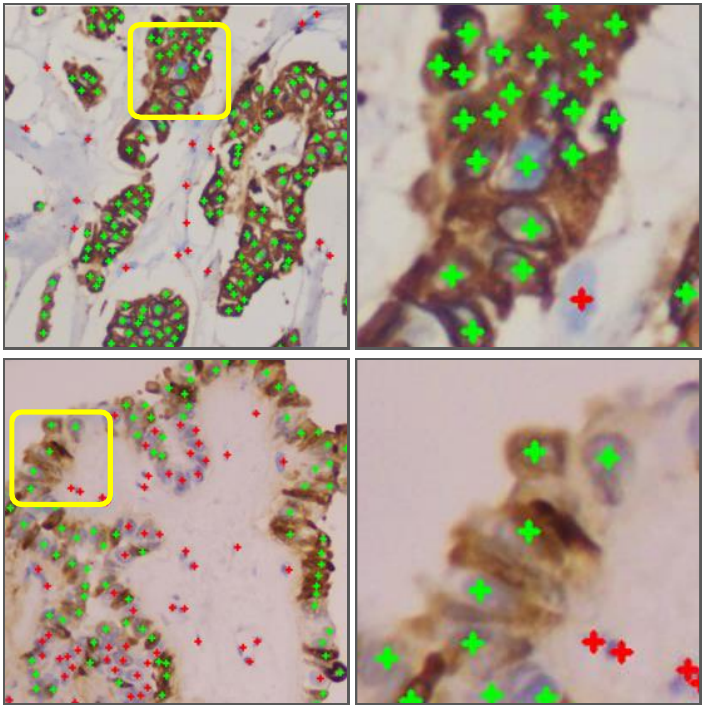}
}
% \vspace{-0.3cm}
\caption{Visualization of ablation experiments. The images in the first row illustrate that the type of some misclassified cells has been corrected by $Ours^{++}$. And the second row shows that the problem of missed detection is alleviated by $Ours^{++}$.}
\label{fig:vs}
\end{figure*}

where $\lambda_c$, $\lambda_p$ and $\lambda_d$ are the weights for $L_c$, $L_p$ and $L_d$ respectively. 
% In the training process, Adam with a momentum 0.9, a weight decay $2\times10^{-5}$ serve as the optimizer and the learning rate is initialized to 0.001. We only use the encoder and main decoder for testing. The main decoder will output two probability maps for the two categories through a sigmoid function. The value in the first and second map is the probability that a pixel predicted to non-positive-tumor and positive-tumor cells respectively. Finally, the number of cells in each category is calculated by picking local peak in each probability map\cite{xie2018efficient}.
In the training stage, Adam serves as the optimizer with the followed setting:  momentum is 0.9, weight decay is $2\times10^{-5}$ and the initial learning rate is 0.001. The hyperparameters $\lambda_c$, $\lambda_p$ and $\lambda_d$ are set to 0.5, 0.5 and 1. While in the testing stage, we only adopt the encoder and main decoder to recognize cells, which will output two probability maps as pixel-level predictions for the two categories through a sigmoid function. Finally, the number of cells in each category is calculated by picking the local peak in probability maps~\cite{xie2018efficient}.

% \vspace{-0.25cm}
\subsection{Results and discussion}
% To evaluate the performance of the proposed multi-task learning framework, and compared with other related works\cite{qu2020weakly,xie2018efficient}, we measured the $total F1$ and the $mean F1$. We get $total F1$ from $recall$ and $precision$ based on all cells in the testing set. And we calculate the F1 score of a testing image, and then average the F1 score of each image in the test set to get the $mean F1$. Therefore, $mean F1$ can effectively reflect the performance in each testing image, especially when the number of positive-tumor cells is relatively small. $(P)$ and $(N)$ indicate positive tumors and non-positive tumors, separately. $Ours$ is the models with encoder and only main decoder. On the basis of $Ours$, $Ours^{+}$ has an auxiliary branch(dynamic decoder) and $Ours^{++}$ has two auxiliary branches(dynamic decoder and prior decoder). Table \ref{tab:res} indicates that the whole framework we proposed get the best results, especially for positive-tumor cells. Moreover, the $mean F1$ of $Ours^{++}$ is higher than the model FCRN models by at least 7\%. So our framework effectively alleviates missed detection of positive tumor cells. 

We use $mean F1$ and $total F1$ to evaluate the performance of the proposed multi-tasking learning framework, where $total F1$ is composed of $recall$ and $precision$ of all cells in the testing set, and $mean F1$ is the average of F1 score of each image in the test set. 
Therefore, $mean F1$ can effectively reflect the performance in each testing image, especially when the number of positive-tumor cells is relatively small. 
The related works~\cite{qu2020weakly,xie2018microscopy} are compared with ours under identical experimental conditions. 
The results are shown in Table~\ref{tab:res},

\begin{table}[!t]
\begin{center}
\vspace{-0.3cm}
\caption[]{Comparison of experimental results. $(P)$ and $(N)$ are positive-tumor cells and the other cells respectively.}
\label{tab:res}
\resizebox{\columnwidth}{!}{
\begin{tabular}{ccccc}
\hline\noalign{\smallskip}
Method& $total F1(N)$& $total F1(P)$& $mean F1(N)$ &$mean F1(P)$\\
\hline\noalign{\smallskip}
$FCRN$\cite{xie2018microscopy} &0.712 &0.719 &0.726 &0.674\\
$EstimateMask$\cite{qu2020weakly} &0.737 &0.777 &0.739 &0.716\\
$Ours$ &0.758 &0.774 &0.759 &0.699\\
$Ours^{+}$  &\textbf{0.759} &0.784 &\textbf{0.763} &0.713\\
$Ours^{++}$  &0.745 &\textbf{0.789} &0.755 &\textbf{0.753}\\
\noalign{\smallskip}\hline
\end{tabular}}
\vspace{-0.3cm}
\end{center}
\end{table}

where $(P)$ and $(N)$ indicate positive tumors and the other cells, respectively. 
In addition, we conduct the ablation experiment to validate the effect of the proposed improvements. In Table~\ref{tab:res}, $Ours$ is the models with encoder and only main decoder. On the basis of $Ours$, $Ours^{+}$ has an auxiliary branch (dynamic decoder) and $Ours^{++}$ has two auxiliary branches (dynamic decoder and prior decoder). 

\begin{figure}
\centering
% \vspace{-0.3cm}
\subfigure[total F1 /P]{
\includegraphics[width=0.47\columnwidth]{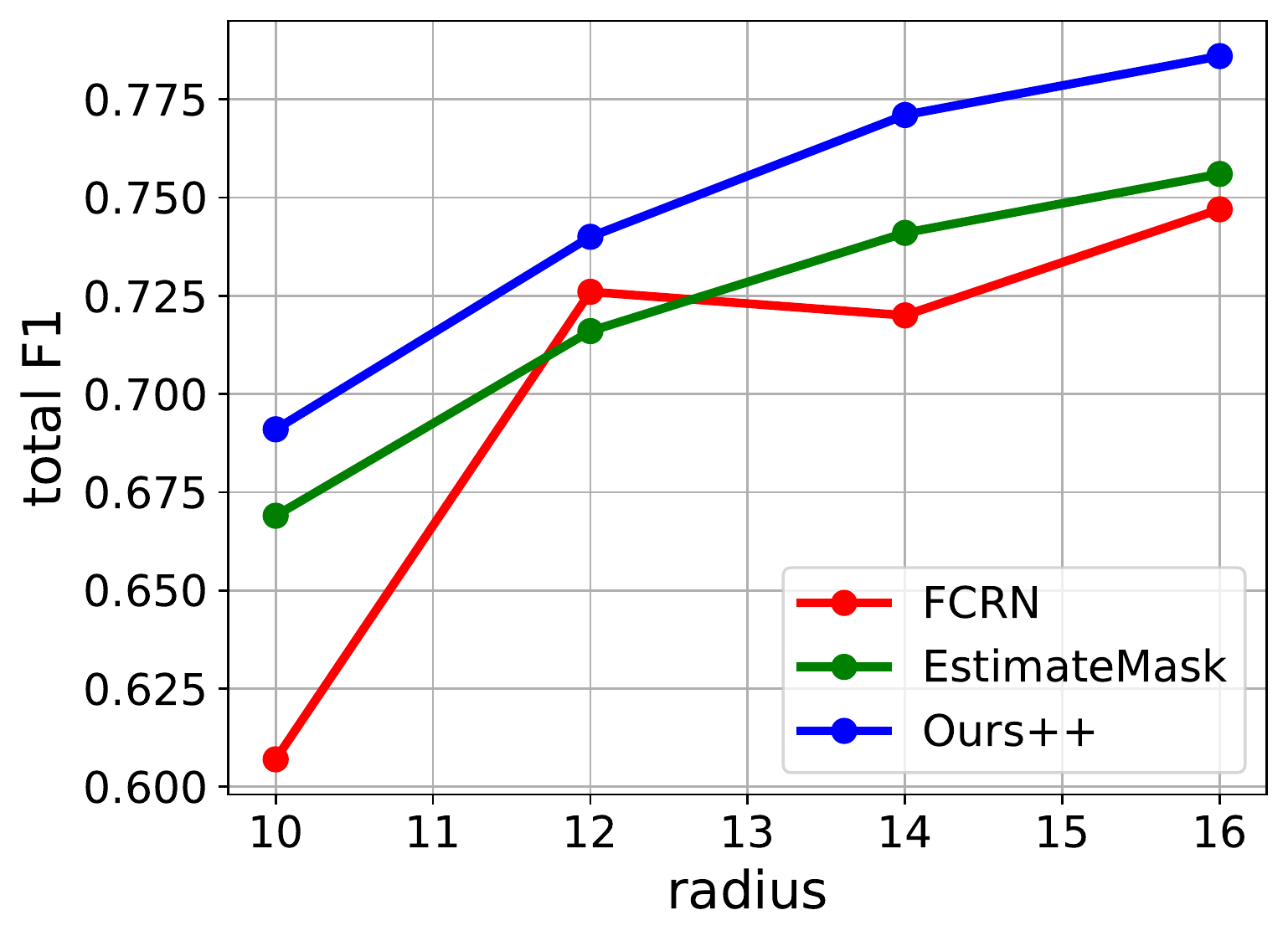}
}
\subfigure[mean F1 /P]{
\includegraphics[width=0.47\columnwidth]{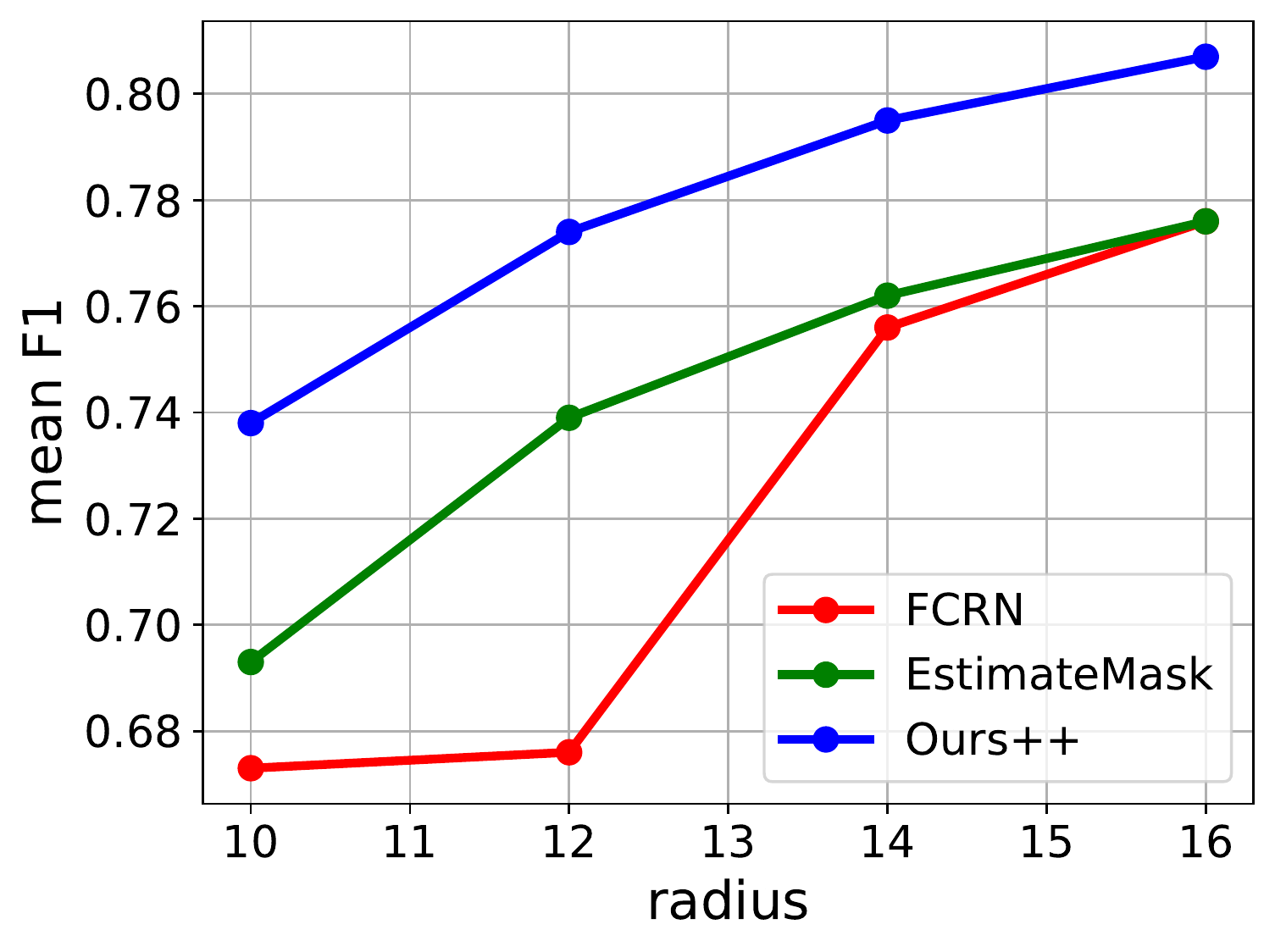}
}
% \vspace{-0.3cm}
\caption{The vertical axis of the two figures are $total F1$ and $mean F1$ of positive-tumor(P) cells respectively, and the horizontal axis is radius $r$, which means the threshold of the distance between the predicted and the annotated location.}
\label{fig:F1_curve}
\end{figure}

The whole framework improves the performance on cell classification as shown in the first row of Fig.\ref{fig:vs} by adding the prior decoder.
In the second row, the model $Ours^{++}$ effectively suppresses the missed detection of positive-tumor cells compared with $Ours$.  Thus, the auxiliary branches will both help the representation learning of the encoder.

Considering that post-processing will affect the statistical results, we tried different radius $r$. We regard the output as a successful prediction if the distance between the center of a predicted cell and the point annotation is less than $r$. The radius $r$ shown in Fig.\ref{fig:F1_curve} represents the number of pixels. Our method has obvious advantages with all the reasonable radius. 

\section{Conclusion}
In this paper, we propose a cell recognition model based on multi-task learning. The main task is for cell classification and counting based on cell segmentation with point annotations. 
% Considering the unique challenges for cell recognition in the IHC cytoplasm staining images, 
We append two auxiliary tasks to provide more information for the encoder. The first auxiliary task (prior decoder) learns the tissue prior information to provide spatial distribution basis apart from shape, size and color for cell classification. And the tissue prior is obtained by another pre-trained model instead of manual labeling. The second auxiliary task (dynamic decoder) is for cell segmentation based on dynamic masks generated from point annotations. Last, the consistency loss between the output of the dynamic decoder and the main decoder is minimized. 
% For some error-prone cells, the auxiliary branch (prior decoder)  provides spatial distribution as an additional cell classification basis. For the missed detection due to differences in cell morphology, the consistency learning between the output of main branch (main decoder) and the auxiliary branch (dynamic decoder) is used to avoid overfitting to the proximity masks. 
% The proposed model is most suitable for cell recognition in the IHC cytoplasm staining images, which is verified in the experiment section. 
The proposed model is more suitable for cell recognition in the IHC cytoplasm staining images by comparing with related works.
A limitation is that this method is sensitive to the hyperparameters in post-processing. And it has become the focus of our follow-up research.

% \begin{figure}[htb]
% \centering
% \includegraphics[width=0.4\columnwidth]{figs/F1.png} 
% \caption{Example of placing a figure with experimental results.}
% \label{fig:F1}
% \end{figure}
% \begin{figure}[htb]
% \centering
% \includegraphics[width=0.4\columnwidth]{figs/mean_F1.png} 
% \caption{Example of placing a figure with experimental results.}
% \label{fig:mean_F1}
% \end{figure}

%proximity
\bibliographystyle{IEEEbib}
\bibliography{strings,refs}

\end{document}